\documentclass[12pt]{iopart}
\usepackage{iopams} 
\usepackage{graphicx}
\usepackage{subfig}
\usepackage{cancel}
\usepackage{soul}
\usepackage{color}
\usepackage{tabularx}
\usepackage{booktabs}
\usepackage{footnote}
\usepackage{threeparttable}  
\usepackage{hyperref}

\begin{document}

\title[Revealing the Source of the Radial Flow Patterns in Proton-Proton Collisions]{Revealing the Source of the Radial Flow Patterns in Proton-Proton Collisions using Hard Probes}

\author{Antonio Ortiz$^{1}$, Gyula Benc\'edi$^{1,3}$, H\'ector Bello$^{1,2}$ }

\address{$^{1}$ Instituto de Ciencias Nucleares, UNAM, M\'exico City, M\'exico}
\address{$^{2}$ Facultad de Ciencias F\'isico Matem\'aticas, BUAP, 1152, Puebla, M\'exico}
\address{$^{3}$ Wigner Research Centre for Physics of the HAS, Budapest, Hungary}
\eads{\mailto{bencedi.gyula@wigner.mta.hu}}

\date{Received: date / Accepted: date}


\begin{abstract}

In this work, we propose a tool to reveal the origin of the collective-like phenomena observed in proton--proton collisions. We exploit the fundamental difference between the underlying mechanisms, 
color reconnection (CR) and hydrodynamics, which produce radial flow patterns in \textsc{Pythia}~8 and \textsc{Epos}~3, respectively. Specifically, we proceed by examining the strength of the coupling 
between the soft and hard components which, by construction, is larger in \textsc{Pythia}~8 than in \textsc{Epos}~3.\\
We study the transverse momentum ($p_{\rm T}$) distributions of charged pions, kaons and (anti)protons in inelastic pp collisions at $\sqrt{s}=$ 7\,TeV produced at mid-rapidity. Specific selections are 
made on an event-by-event basis as a function of the charged particle multiplicity and the transverse momentum of the leading jet ($p_{\rm T}^{\rm jet}$) reconstructed using the \textsc{FastJet} algorithm at 
mid-pseudorapidity ($|\eta|<1$).
From our studies, quantitative and qualitative differences between \textsc{Pythia}~8 and \textsc{Epos}~3 are found in the $p_{\rm T}$ spectra when (for a given multiplicity class) the leading jet $p_{\rm T}$ 
is increased. In addition, we show that for low-multiplicity events the presence of jets can produce radial flow-like behavior. Motivated by our findings, we propose to perform a similar analysis using 
experimental data from RHIC and LHC.

\end{abstract}


\maketitle


\section{Introduction}

The study of particle production in high-multiplicity events in small collision systems at the LHC has revealed unexpected new collective-like phenomena. In particular, for high-multiplicity proton--proton (pp) and 
proton--lead (p--Pb) collisions, radial flow signals~\cite{Abelev:2013haa,Adam:2016dau}, long-range angular correlations~\cite{ABELEV:2013wsa,Khachatryan:2016txc}, and the strangeness enhancement~\cite{Adam:2016emw,Adam:2015vsf,Khachatryan:2016yru} 
have been reported. Those effects are well-known in heavy-ion collisions, where they are attributed to the existence of the strongly interacting Quark-Gluon Plasma (QGP)~\cite{Bala:2016hlf,Abelev:2014laa,Adam:2015kca}. 
Understanding the phenomena is crucial because, for heavy-ion physics, pp and p--Pb collisions have been used as the baseline (``vacuum'') to extract the genuine QGP effects. However, it is worth mentioning that no jet 
quenching effects have been found so far in p--Pb collisions~\cite{Adam:2015hoa}, suggesting that other mechanisms could also play a role in producing collective-like behavior in small collision systems~\cite{Ortiz:2015cma,Zakharov:2015gza}.

Hydrodynamic calculations reproduce many of the observations qualitatively~\cite{Bozek:2013ska}. However, it has also been found that multi-parton interactions (MPI) ~\cite{Sjostrand:1987su} and color reconnection (CR) 
as implemented in \textsc{Pythia}~\cite{Sjostrand:2014zea} produce radial flow patterns via boosted color strings~\cite{Ortiz:2013yxa}. Moreover, within the dilute-dense limit of the color glass condensate, it has been 
demonstrated that the physics of fluctuating color fields can generate azimuthal multi-particle correlations~\cite{Lappi:2015vta};  and the mass ordering of elliptic flow when the fragmentation implemented in \textsc{Pythia} 
is included~\cite{Schenke:2016lrs}. The same observable has been studied using the multi-phase transport model~\cite{Ma:2014pva}, where the ridge structure can be generated assuming incoherent elastic scatterings of partons 
and the string melting mechanism. Other mechanisms like ``color ropes'', which are formed by the fusion of color strings close in space, can increase both the strangeness production and the radial flow-like effects~\cite{Bierlich:2014xba}.

The measurements of the transverse momentum ($p_{\rm T}$) spectra of identified particles as a function of event multiplicity in pp collisions at the LHC~\cite{Adam:2016emw,Chatrchyan:2012qb} have shown that models fail to 
describe the data quantitatively. Therefore, the results of those comparisons alone are not enough to give desired information about the origin of the observed effects (i.e. radial flow-like patterns). In order to extract 
more information, we propose the implementation of a differential study based on the classification of the events according to event multiplicity and the jet content.

In the so-called MPI-based model of color reconnection~\cite{Sjostrand:2014zea}, the interaction between scattered partons at soft and at hard $p_{\rm T}$ scales is imposed as follows. All gluons of low-$p_{\rm T}$ interactions 
can be inserted onto the colour-flow dipoles of a higher-$p_{\rm T}$ one, keeping the total string length as short as possible. Since the probability of having a hard scattering increases with the number of MPI, color reconnection 
can give a strong correlation between the radial flow-like patterns and the hard component of the collision~\cite{Cuautle:2015kra} in high-multiplicity events. 

On the contrary, in the scenario where the hydrodynamical evolution of the system is the prime mechanism, jets are not expected to strongly modify the radial flow patterns. Albeit hard partons cannot thermalize, momentum loss of 
jets could affect the fluid dynamic evolution of the medium. However, the effect has been studied for heavy-ion collisions and it was found to give only a minor correction~\cite{Floerchinger:2014yqa}. In the present paper we argue 
that by exploiting such a fundamental difference between both models, one might say whether or not the observed effects are driven by hydrodynamics. To this end, we propose a systematic study by analyzing the mid-rapidity ($|y|<1$) 
inclusive $p_{\rm T}$ spectra of identified charged hadrons as a function of the  mid-pseudorapidity ($|\eta|<1$) event multiplicity ($N_{\rm ch}$) and transverse momentum of the leading jet ($p_{\rm{T}}^{\rm jet}$). This study was 
carried out using \textsc{Pythia}~8.212 and \textsc{Epos}~3.117 Monte Carlo (MC) event generators, from now on referred to as \textsc{Pythia}~8 and \textsc{Epos}~3, respectively.\\

The paper is organized as follows: In section 2, some important features of the Monte Carlo event generators and the jet finder are outlined, with special emphasis on those aspects which are relevant in this study. In section 3, the 
results and discussion are presented. Finally, in section 4, the conclusions are given.


\section{Simulation setup and Monte Carlo models}

\begin{figure*}
  \includegraphics[width=0.93\textwidth]{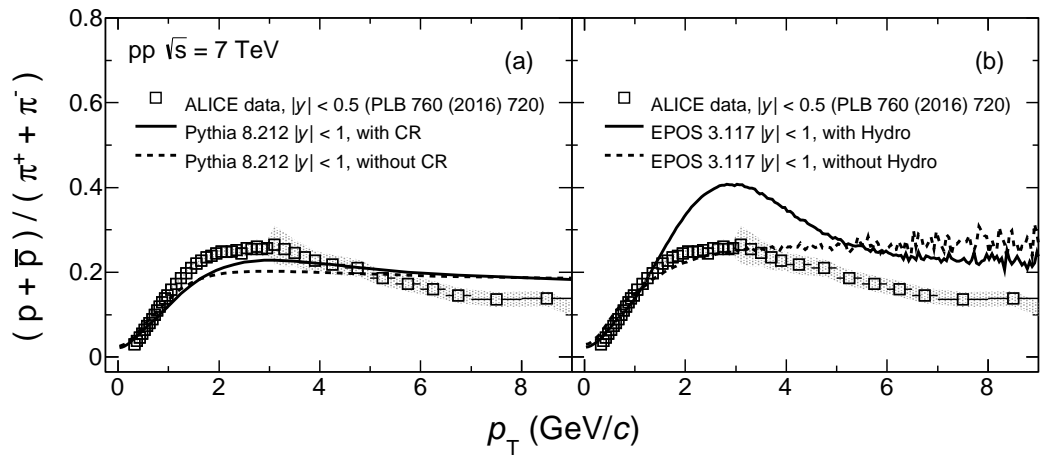}     
  \caption{\small{Proton-to-pion ratio as a function of $p_{\rm T}$ for inelastic pp collisions at $\sqrt{s}=$ 7\,TeV measured by the ALICE Collaboration~\cite{Adam:2016dau}. 
  Results are compared to models carried out with (a) \textsc{Pythia}~8 and (b) \textsc{Epos}~3 event generators. Cases with and without the effect of color reconnection and 
  hydrodynamics are plotted as solid and dashed lines, respectively}}
  \label{fig:ptopi:epos:inel}
\end{figure*}

The studies were carried out for pp collisions at the centre-of-mass energy of $\sqrt{s}=$ 7\,TeV considering sets with and without the mechanism which produces radial flow patterns. To illustrate how \textsc{Pythia}~8 and \textsc{Epos}~3 
describe the experimental data measured by the ALICE Collaboration~\cite{Adam:2016dau}, Fig.~\ref{fig:ptopi:epos:inel} shows the proton-to-pion ratio for inelastic pp collisions at $\sqrt{s}=$ 7\,TeV. In the case of \textsc{Pythia}~8 shown 
in Fig.~\ref{fig:ptopi:epos:inel}(a), as discussed in~\cite{Ortiz:2013yxa}, the model shows a qualitative agreement with data, e.g., the bump in the proton-to-pion ratio, though the size of the effect is underestimated. 
The same level of accuracy is achieved by \textsc{Epos}~3 which according to Fig.~\ref{fig:ptopi:epos:inel}(b) overestimates the effect when hydrodynamics is included. So, in the end, we will compare models which still do not fully describe 
the data, but this is beside the point of our interest here. Instead, we shall study differences attributed to the fundamental underlying physics mechanisms which produce the observed radial flow effects.

The results are presented both for \textsc{Pythia}~8 and \textsc{Epos}~3 using primary charged particles, defined as all charged particles produced in the collision including the products of strong and electromagnetic decays but excluding 
products of weak decays. Unless stated otherwise, no requirement on the minimum $p_{\rm T}$ of the particles is applied in any of the results. All MC generators use parton-to-hadron fragmentation approaches fitted to the experimental 
data\,--\,such as the Lund string~\cite{Andersson:1983ia} and area law hadronization~\cite{Artru:1974hr} models.\\
For the results presented in this paper, we generated $\approx10^2$ million inelastic events, including diffractive and non-diffractive components. A non-diffractive event includes a parton-parton interaction with a large momentum transfer 
(more than a few GeV/$c$). On the other hand, some inelastic collisions can be ``diffractive'', in which a virtual particle (known as Pomeron) is responsible for the interaction. This sample was subsequently split into sub-samples based on 
the selection of charged particle multiplicity measured at mid-pseudorapidity and on the hardness of the event by imposing a cut on the jet transverse momentum ($p_{\rm{T}}^{\rm jet}$) found within the same acceptance.

  \subsection{\textsc{Epos}~3 and hydrodynamics}

\textsc{Epos}~3~\cite{Werner:2013tya,Werner:2010ny} is a generator of complete events (soft + hard components) which contains hard scatterings and MPI. For high string densities, e.g. those achieved in high-multiplicity pp collisions, 
the model does not allow the strings to decay independently, instead, if energy density from string segments is high enough they fuse into the so-called ``core'' region, which evolves hydrodynamically. On the other hand, the low 
energy density region forms the ``corona'' which hadronizes using the unmodified string fragmentation. 


The ``core'' region originates around 30\,\% of the central particle production for an average pp collision at $\sqrt{s}=$ 7\,TeV, $\langle {\rm d}N_{\rm ch}/{\rm d}\eta \rangle_{|\eta|<2.4} \approx 6.25$. This fraction might reach  
$\approx 75$\,\% for $\langle {\rm d}N_{\rm ch}/{\rm d}\eta \rangle_{|\eta|<2.4} \approx 20.8$~\cite{Martin:2016igp}. Concerning the hard component, the inclusive jet cross section for pp collisions at $\sqrt{s}=$ 0.2\,TeV obtained 
with \textsc{Epos}~3 agrees within 5\,\% and 4\,\% with STAR data and NLO pQCD calculations, respectively~\cite{Porteboeuf:2010um}. Therefore, an analysis in \textsc{Epos}~3 as a function of event multiplicity and leading jet transverse 
momentum is suitable to our needs.

\begin{figure*}
  \centering
  \includegraphics[width=0.93\textwidth]{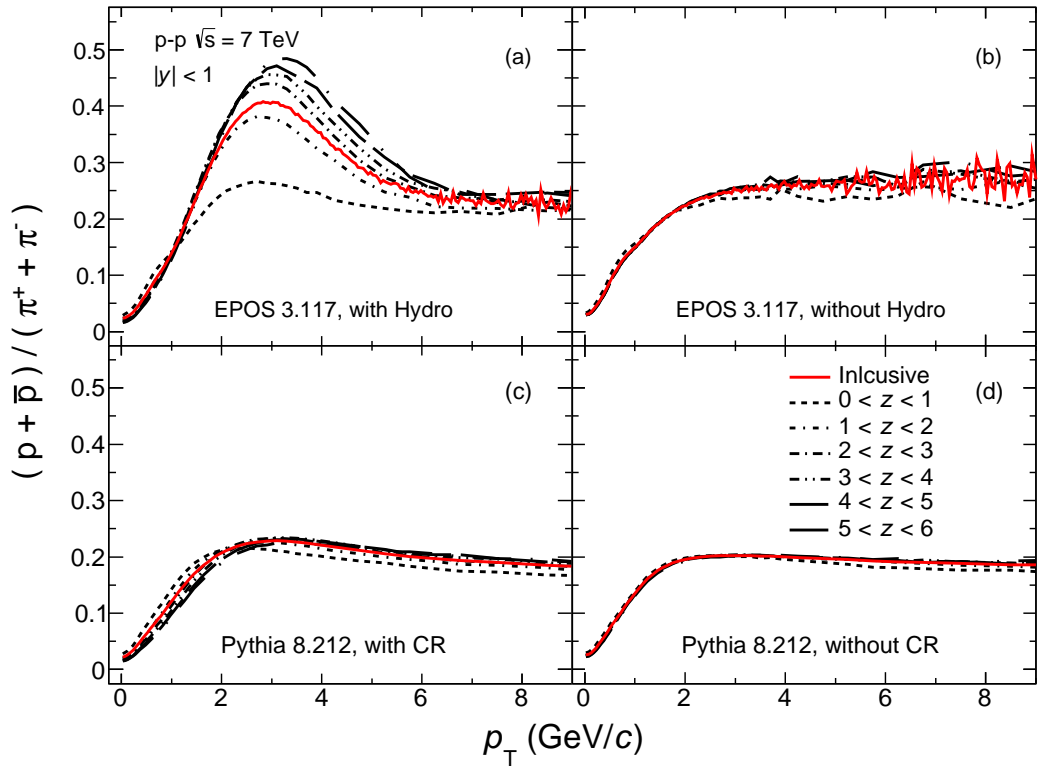}     
  \caption{(Color online) \small{Proton-to-pion ratio as a function of $p_{\rm T}$ for different multiplicity event classes. Results for pp collisions at $\sqrt{s}=$ 7\,TeV 
  generated with \textsc{Epos}~3 and \textsc{Pythia}~8 are presented. For \textsc{Pythia}~8 (\textsc{Epos}~3) the ratios are displayed for simulations with and without color 
  reconnection (hydrodynamical evolution of the system).}}
  \label{fig:ptopi:epospy}
\end{figure*}

To illustrate the effect of hydrodynamics on flow observables, Fig.~\ref{fig:ptopi:epospy} shows the proton-to-pion ratio as a function of $p_{\rm T}$ for different multiplicity classes, including also the inclusive case as a reference. 
Results are presented in intervals of $z$, defined as
\begin{equation}\label{eq:zed}
z=\frac{ {\rm d}N_{\rm ch}/{\rm d}\eta }{ \langle {\rm d}N_{\rm ch}/{\rm d}\eta \rangle }~,
\end{equation}
where $\langle {\rm d}N_{\rm ch}/{\rm d}\eta \rangle~(=5.505)$ is the average mid-rapidity particle density for inelastic pp collisions at $\sqrt{s}=$ 7\,TeV. It is important to mention that according to ALICE results~\cite{Adam:2016emw} 
$\langle {\rm d}N_{\rm ch}/{\rm d}\eta \rangle \sim 25$ is already large enough to see new phenomena in small systems, like the strangeness enhancement. For the simulations presented in this paper, such high multiplicity densities are achieved for $z>4$. 
Fig.~\ref{fig:ptopi:epospy}(a) shows the case when hydrodynamics is considered in the simulations. In this case, a clear evolution of the particle ratio with $z$ can be observed, i.e. going from the lowest $z$ to the highest $z$ values the particle 
ratios exhibit a depletion (enhancement) for $p_{\rm T}<1$\,GeV/$c$ ($1<p_{\rm T}<6$\,GeV/$c$). This feature is usually attributed to radial flow which modifies the spectral shapes of the $p_{\rm T}$ distribution depending on the hadron mass. 
On the contrary, Fig.~\ref{fig:ptopi:epospy}(b) shows the case without hydrodynamics, where the particle ratios do not evolve with multiplicity.

  \subsection{\textsc{Pythia}~8 and color reconnection}

\textsc{Pythia}~8~\cite{Sjostrand:2014zea} is a full event generator for pp collisions. For inelastic collisions, which is the main interest here, each collision is modeled via one or more parton-parton interactions. The full calculation 
involves leading-order (LO) pQCD $2\rightarrow 2$ matrix elements, complemented with initial- and final-state parton radiation, multiple particle interactions, beam remnants and the Lund string fragmentation model. \textsc{Pythia}~8 also 
has strong final-state parton interactions (implemented through the CR models~\cite{Sjostrand:2014zea}). In this work, we use the Monash 2013 tune~\cite{Skands:2014pea} which has as default parametrisation the so-called MPI-based model of 
color reconnection. Such a model allows partons of each MPI system to form their own structure in color space and then, they are merged into the color structure of a higher $p_{\rm T}$ MPI system, with a probability $\mathcal{P}$ given by:
\begin{equation}
\mathcal{P}(p_{\rm T})=\frac{(R\times p_{\rm T0})^{2}}{(R\times p_{\rm T0})^{2}+p_{\rm T}^{2}}~,
\end{equation}
where $R$ is the reconnection range ($0\leq R\leq 10$) and $p_{\rm T0}$ is the energy-dependent parameter used to damp the low-$p_{\rm T}$ divergence of the $2\rightarrow 2$ QCD cross section.

Shown in the bottom row of Fig.~\ref{fig:ptopi:epospy} ((c) and (d) panels) are the proton-to-pion ratios for several $z$ classes, with and without CR effects. It is worth noticing that compared to \textsc{Epos}~3 (top row of Fig.~\ref{fig:ptopi:epospy}) 
a similar indication for the radial flow effect is present in \textsc{Pythia}~8, but in this case, the radial flow-like behavior is attributed to color reconnection~\cite{Ortiz:2013yxa}. Also, it is remarkable that, in good approximation, with the 
CR option the magnitude of the peak is constant as a function of $z$.


  \subsection{\textsc{FastJet} 3.1.3 and hardness of the event}
  
Besides the multiplicity selection, we also classify events based on the transverse momentum ($p_{\rm T}^{\rm jet}$) of the produced jets. Per event, jets are reconstructed with the well-known anti-$k_{\rm T}$ algorithm implemented in 
\textsc{FastJet}~3.1.3~\cite{Cacciari:2011ma}, using charged and neutral particles, considering a cone radius of $0.4$ and a minimum transverse momentum $p_{\rm T,min}^{\rm jet}=5$\,GeV/$c$. The lower requirement on the jet $p_{\rm T}$ 
acts to suppress soft interactions by ensuring that at least one semi-hard scattering is present within the acceptance. In the following, the jet searching is done within a given pseudorapidity interval, which defines the maximum 
pseudorapidity of the jet. It is important to highlight that \textsc{FastJet} is a well-known tool for jet reconstruction in heavy-ion collisions, where it has been extensively used, even in most central Pb--Pb collisions. 

\section{Results and discussion}


  \subsection{Multiplicity dependence of the leading jet $p_{\rm T}$}

By running \textsc{FastJet} (considering cone radius $0.4$, $|\eta|<1$ and $p_{\rm T,min}^{\rm jet}=5$\,GeV/$c$) over the sample generated with \textsc{Pythia}~8 we obtain the results for the average mid-pseudorapidity densities, 
presented in Table~\ref{tab:1}. \\
Going from $\langle {\rm d}N_{\rm ch}/{\rm d}\eta \rangle=2.12$ to $\langle {\rm d}N_{\rm ch}/{\rm d}\eta \rangle=46.1$, the average leading jet $p_{\rm T}$ ranges from $7.09$\,GeV/$c$ up to $19.7$\,GeV/$c$. A similar behavior was 
found for the leading parton transverse momentum, obtained at mid-pseudorapidity, as a function of $\langle {\rm d}N_{\rm ch}/{\rm d}\eta \rangle$. The effect is explained in the context of multi-partonic interactions because the 
probability of finding a hard parton is expected to be larger in high-multiplicity events (large average $N_{\rm mpi}$) than in low-multiplicity events (small average $N_{\rm mpi}$).  This effect is also reflected 
in the behavior of the fraction of events having at least one jet with momentum above $5$\,GeV/$c$. In the second column of Tab~\ref{tab:1} we included the corresponding event multiplicity class defined by Eq.~\ref{eq:zed} in comparison 
to the other measured variables. For the presented results we considered events having $z$ from the lowest possible value ($0<z<1$) up to $5<z<6$. One can see that for the highest considered $z$ interval ($5<z<6$) more than $90$\,\% of 
the events contain jets with $p_{\rm T}^{\rm jet}>5$\,GeV/$c$. This feature is simply the result of the selection bias, i.e. the higher the multiplicity at mid-pseudorapidity the higher the probability to find jets within the same $\eta$ region.

\begin{table}
\centering
\begin{tabular}{cccc}
\hline\noalign{\smallskip}
$\left\langle \frac{{\rm d}N_{\rm ch}}{{\rm d}\eta} \right\rangle_{|\eta|<1}$ & $\frac{\mathrm{d}\,N_{\rm ch}/\mathrm{d}\,\eta}{\left\langle\mathrm{d}\,N_{\rm ch}/\mathrm{d}\,\eta\right\rangle_{|\eta|<1}}~(\equiv z)$  & $\langle p_{\rm T}^{\rm jet} \rangle_{|\eta|<1}$ (GeV/$c$) & \% of events with   \\
 &  &  & $p_{\rm T}^{\rm jet}>5$\,GeV/$c$   \\
\noalign{\smallskip}\hline\noalign{\smallskip}
2.12 & $0<z<1$  &   7.09 &   1.03 	\\
8.12 & $1<z<2$  &   7.49 &  13.1 	\\
13.6 & $2<z<3$  &   7.83 &  37.3 	\\
19.0 & $3<z<4$  &   8.48 &  63.7 	\\
24.4 & $4<z<5$  &   9.56 &  83.2 	\\
29.8 & $5<z<6$  &  11.1  &  93.9 	 \\
35.2 & $6<z<7$  &  13.2  &  98.2 	 \\
40.6 & $7<z<8$  &  16.1  &  99.5 	 \\
46.1 & $8<z<9$  &  19.7  &  99.8 	 \\
\noalign{\smallskip}\hline
\end{tabular}
\caption{\small{Charged-particle pseudorapidity densities at central pseudorapidity ($|\eta|<1$), for pp collisions at $\sqrt{s}=$ 7\,TeV simulated with 
\textsc{Pythia}~8 having jets with $p_{\rm T}$ above $5$\,GeV/$c$ in the same pseudorapidity region. The multiplicity classes are presented along the leading jet $p_{\rm T}$ 
and the fraction of events where a jet with $p_{\rm T}$ above $5$\,GeV/$c$ was identified.}}
\label{tab:1}
\end{table}


  \subsection{Proton-to-pion ratio as a function of $z$ and $p_{\rm T}^{\rm jet}$}

\begin{figure*}
  \centering
  \includegraphics[width=0.93\textwidth]{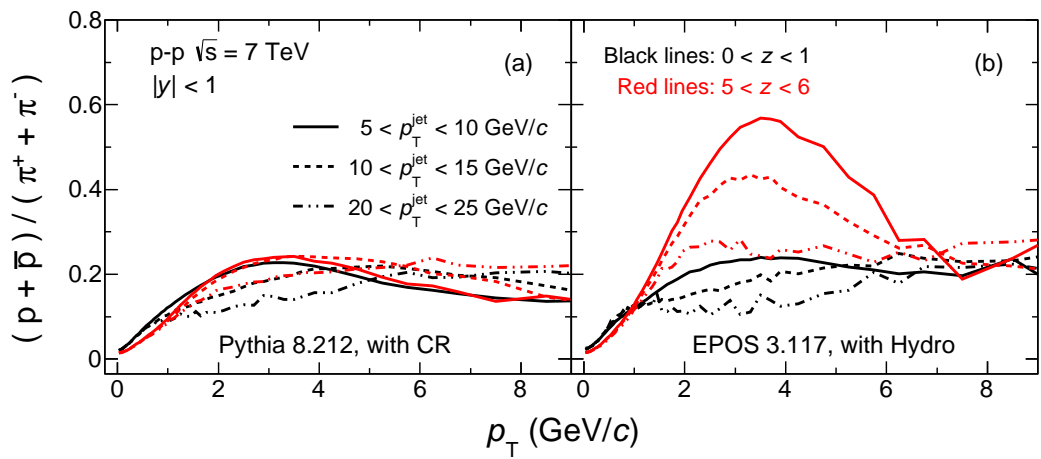}     
  \caption{(Color online) \small{Proton-to-pion ratio as a function of $p_{\rm T}$ for two multiplicity classes, 
  $0<z<1$ (black lines) and $5<z<6$ (red lines); and for different $p_{\rm T}^{\rm jet}$ intervals. Results are shown for 
  both (a) \textsc{Pythia}~8 and (b) \textsc{Epos}~3.}}
  \label{fig:fastjet:lowz}
\end{figure*}

Figure~\ref{fig:fastjet:lowz} shows the proton-to-pion ratio as a function of $p_{\rm T}$ for event classes with low and high $z$. Regarding the low-$z$ case ($0<z<1$), results indicate that for $5<p_{\rm T}^{\rm jet}<10$\,GeV/$c$ the 
ratios exhibit an enhancement at $p_{\rm T}\approx 3$\,GeV/$c$ for both \textsc{Epos}~3 and \textsc{Pythia}~8. If the $p_{\rm T}^{\rm jet}$ is increased then the position of the observed peak is shifted to higher $p_{\rm T}$. This observation 
suggests that the peak is not an exclusive effect of radial flow (as suggested by Fig.~\ref{fig:ptopi:epospy}), but also a feature of the fragmentation. It is worth noticing that the same effect has been observed in ALICE data, where the jet 
hadrochemistry has been measured in minimum bias pp collisions at $\sqrt{s}=$ 7\,TeV~\cite{Lu:2014roa}.
  
    
Also shown in Fig.~\ref{fig:fastjet:lowz} the multiplicity class $5<z<6$, where we see that the maximum of the proton-to-pion ratio increases with increasing multiplicity, showing significantly stronger behavior for \textsc{Epos}~3 than for 
\textsc{Pythia}~8. In \textsc{Epos}~3, the event class $5<p_{\rm T}^{\rm jet}<10$\,GeV/$c$ exhibits an enhancement of the $({\rm p+ \bar{p}})/(\pi^{+}+\pi^{-})$ ratio with respect to the inclusive case shown in Fig.~\ref{fig:ptopi:epospy} 
without any selection on $p_{\rm T}^{\rm jet}$. Going to higher $p_{\rm T}^{\rm jet}$, the position of the peak is shifted to lower $p_{\rm T}$ values and the magnitude of the peak is significantly smaller than in the inclusive case. 
In contrast, the effect is quantitatively and qualitatively different in \textsc{Pythia}~8. Namely, the height of the peak is approximately independent of the increase of $p_{\rm T}^{\rm jet}$, instead, there is a moderate shift towards 
higher $p_{\rm T}$. In \textsc{Epos}~3, the effect vanishes when hydrodynamical effects are switched off\,--\,in a similar way as seen in Fig.~\ref{fig:ptopi:epospy}(b). Therefore, it can be a consequence of the ``core-corona'' separation, 
where low-momentum partons are more likely forming the ``core'' region. It is worth mentioning that this difference between the two event classes could contribute to the differences observed in the hadrochemistry measured in the so-called 
``bulk'' (outside the jet peak) and the jet regions in p--Pb and Pb--Pb collisions at the LHC~\cite{Veldhoen:2012ge,XZhang:2014sva}. To summarize, an analysis of data as a function of event multiplicity and the hardness of the event provides 
a more powerful tool for testing the aforementioned models than the one which considers selection based only on event mutliplicity.

  \subsection{Blast-wave model fits}

Color reconnection, without any hydrodynamical component, produces radial flow-like patterns in events simulated with \textsc{Pythia}~8, as shown in Ref.~\cite{Ortiz:2013yxa}. Such a conclusion was based on the good agreement between the 
Boltzmann--Gibbs blast-wave model and the $p_{\rm T}$ spectra of different particle species. The blast-wave model describes a locally thermalised medium which experiences a collective expansion with a common velocity field and undergoing an 
instantaneous common freeze-out~\cite{Schnedermann:1993ws}. 
The functional form of the model is given by:
\begin{equation}
\frac{\mathrm{d}N}{p_{\rm T}\,\mathrm{d}p_{\rm T}} \propto \int_{0}^{R}\,r\,\mathrm{d}r\,m_{\rm T}\,I_{0}\left(\frac{p_{\rm T}\sinh\rho}{T_{\rm kin}}\right)\,K_{1}\left(\frac{m_{\rm T}\cosh\rho}{T_{\rm kin}}\right)~,
\end{equation}
where $K_{1}$ and $I_{0}$ are the modified Bessel functions, and $T_{\rm kin}$ is the kinetic freeze-out temperature. On the other hand,  $\rho=\tanh^{-1}(\beta_{\rm T})$,  where $\beta_{T}$ is the radial profile of the transverse expansion velocity.\\
From the simultaneous fit of the blast-wave model to the $p_{\rm T}$ spectra of different particle species we extract two parameters: $T_{\rm kin}$ and $\langle \beta_{\rm T} \rangle$. In the current study we considered the $p_{\rm T}$ ranges: $0.5<p_{\rm T}<1.0$\,GeV/$c$, 
$0.3<p_{\rm T}<1.5$\,GeV/$c$ and $0.8<p_{\rm T}<2.0$\,GeV/$c$ to fit the model to the $p_{\rm T}$ distributions of charged pions, kaons and (anti)protons, respectively. The specific selection of the $p_{\rm T}$ ranges mentioned above was 
successfully applied in our previous studies~\cite{Cuautle:2015kra} where the parametrizations, obtained from the fits, described within 10\,\% the strange and multi-strange hadron $p_{\rm T}$ spectra. 

\begin{figure*}
  \centering
  \includegraphics[width=0.9\textwidth]{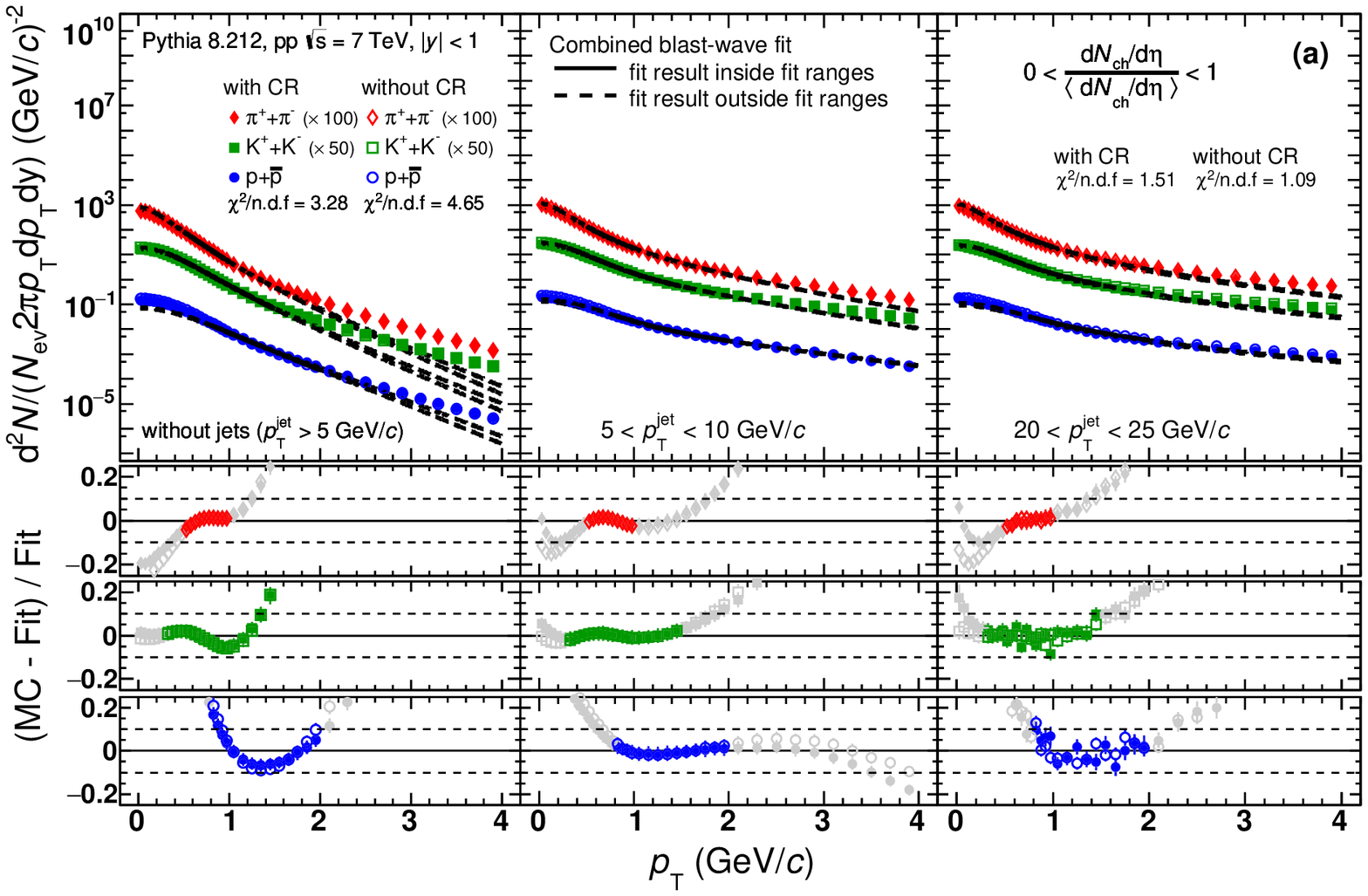}
  \includegraphics[width=0.9\textwidth]{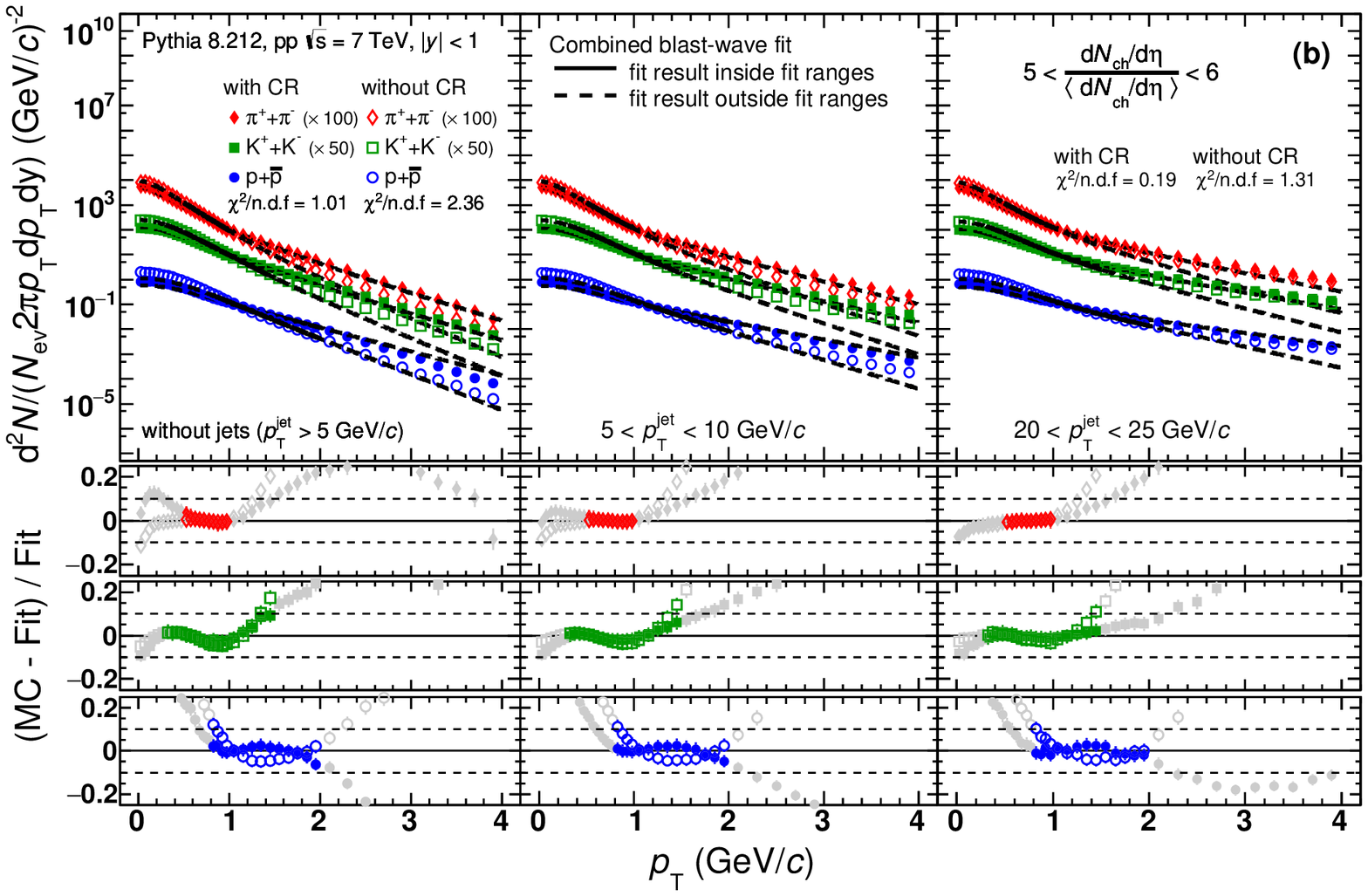}
\caption{(Color online) \small{Transverse momentum distributions of charged pions, kaons and (anti)protons for (a) low- and (b) high-multiplicity pp collisions at $\sqrt{s}=$ 7\,TeV 
generated with \textsc{Pythia}~8. For each multiplicity class, three sub-samples are shown, from left to right, events without a leading jet with $p_{\rm T}^{\rm jet}>5$\,GeV/$c$, 
with $5<p_{\rm T}^{\rm jet}<10$\,GeV/$c$ and with $20<p_{\rm T}^{\rm jet}<25$\,GeV/$c$, respectively. Results for the cases with and without color reconnection (CR) are plotted with 
full and empty markers, respectively. The parametrizations obtained from the simultaneous blast-wave fits are shown as solid lines. The deviation of the $p_{\rm T}$ 
spectrum from the blas-wave model is shown in the bottom panels, indicating the fitting range using colored markers.}}
\label{fig:bwPythia8}
\end{figure*}    

For pp collisions at $\sqrt{s}=$ 7\,TeV simulated with \textsc{Pythia}~8, Fig.~\ref{fig:bwPythia8} shows the $p_{\rm T}$ spectra of charged pions, kaons and (anti)protons for two multiplicity classes, $0<z<1$ (upper panel (a)) and $5<z<6$ 
(lower panel (b)), being each one split into three specific subclasses based on the selection of the jet transverse momentum, $p_{\rm T}^{\rm jet}$. The first subclass is treated as a baseline since no jets at mid-pseudorapidity are found; it is 
compared with samples in which low (5--10\,GeV/$c$) and high (20--25\,GeV/$c$) $p_{\rm T}$ jets are produced. For a given average multiplicity $z$ class and jet transverse momentum $p_{\rm T}^{\rm jet}$ interval two cases are considered: with 
and without color reconnection. \\
There are some interesting observations one might read off from the results of this analysis. Firstly, we concentrate on results obtained from \textsc{Pythia}~8 simulations, since we already know we might expect the presence of radial flow patterns. 
Then we shall be examining the  results  produced by \textsc{Epos}~3.\\
Even at extremely low multiplicity ($0<z<1$), where color reconnection effects are negligible, it is possible to find an event class where the radial flow-like patterns pop up. Especially, 
in events having $p_{\rm T}^{\rm jet}>5$\,GeV/$c$ the $p_{\rm T}$ distributions of identified hadrons are better described by the blast-wave model than in those without jets. Looking at differences between the model fit and the MC result in terms 
of ratios (shown in the bottom rows of Fig.~\ref{fig:bwPythia8}(a)) one can see a smoothening trend with $p_{\rm T}$ when going towards higher $p_{\rm T}^{\rm jet}$ which is hardly visible for pions but more pronounced for kaons and (anti)protons. 
This behavior is also supported by the reduction on the reduced $\chi^{2}$ values: fit results give $\chi^{2}/\mathrm{n.d.f} = 3.28$ ($4.65$) for events without jets and $\chi^{2}/\mathrm{n.d.f} = 1.51$ ($1.09$) for events including jets with 
$20<p_{\rm T}^{\rm jet}<25$\,GeV/$c$ for the case with (without) color reconnection, respectively.\\
It is worth mentioning that recently the CMS Collaboration has reported that in low-multiplicity pp events, the elliptic flow Fourier harmonic is not zero~\cite{Khachatryan:2016txc}, supporting the idea that other mechanisms could produce 
the collective-like behavior.\\
In high-multiplicity events ($5<z<6$), again, examining the relative differences between the spectra and the fits, the overall agreement between the spectra and the blast-wave fits shows a smaller 
dependence with $p_{\rm T}^{\rm jet}$. The blast-wave model fits give a slightly worse description of the spectra when CR is not included in the simulation\,--\,even if a jet with $p_{\rm T}^{\rm jet}>5$\,GeV/$c$ is produced at mid-rapidity\,--\,
than for the case with the inclusion of CR. The ratios of spectra (simulated with CR) to blast-wave model show a modest smoothening trend with increasing $p_{\rm T}^{\rm jet}$ while those without CR stay approximately $p_{\rm T}^{\rm jet}$-independent. 
These observations are confirmed by the fit qualities in terms of reduced $\chi^{2}$ values: $\chi^{2}/\mathrm{n.d.f} = 1.01$ ($2.36$) for events without jets and $\chi^{2}/\mathrm{n.d.f} = 0.19$ ($1.31$) for events including jets 
($20<p_{\rm T}^{\rm jet}<25$\,GeV/$c$) for the case with (without) color reconnection. The observed effects just reflect the fact that in \textsc{Pythia}~8 the interaction between jets and the underlying event is crucial for generating a collective-like behavior.  

\begin{figure*}
  \centering
  \includegraphics[width=0.9\textwidth]{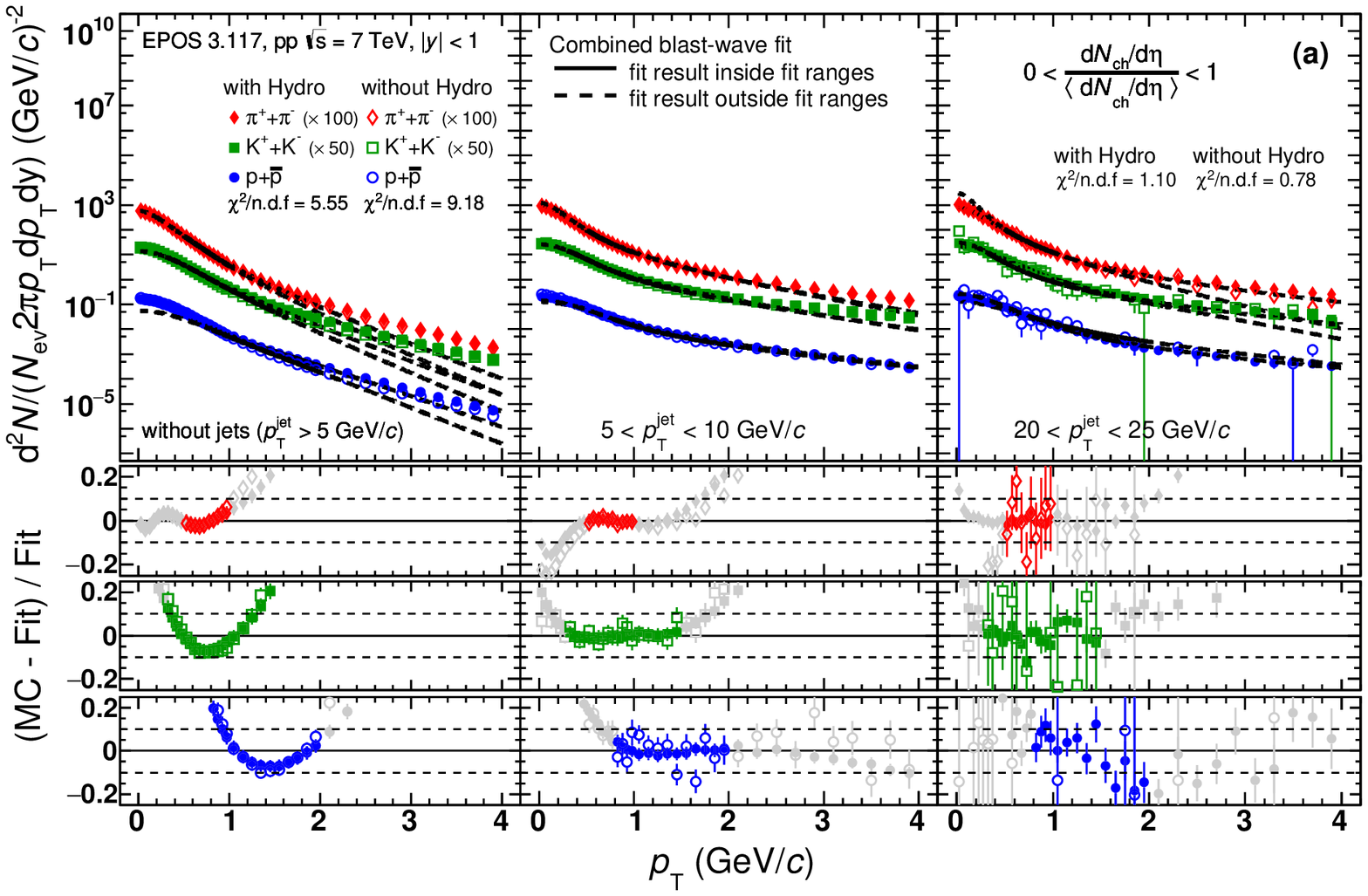}
  \includegraphics[width=0.9\textwidth]{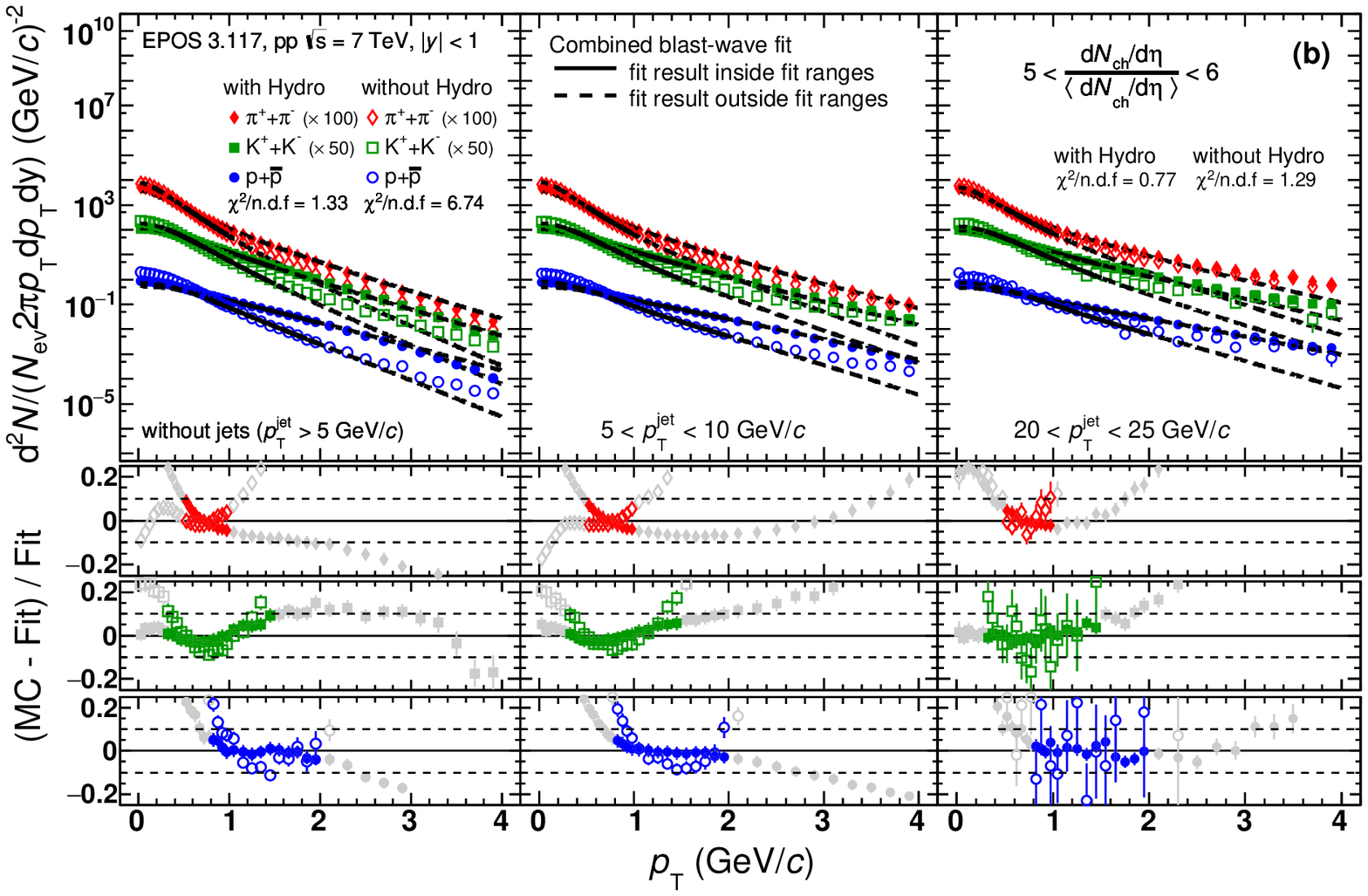}
\caption{(Color online) \small{Transverse momentum distributions of charged pions, kaons and (anti)protons for (a) low- and (b) high-multiplicity pp collisions at $\sqrt{s}=$ 7\,TeV 
generated with \textsc{Epos}~3. For each multiplicity class, three sub-samples are shown. From left to right, events without a leading jet with $p_{\rm T}^{\rm jet}>5$\,GeV/$c$, 
with $5<p_{\rm T}^{\rm jet}<10$\,GeV/$c$ and with $20<p_{\rm T}^{\rm jet}<25$\,GeV/$c$, respectively. Results for the cases with and without hydrodynamics (hydro) are plotted with 
full and empty markers, respectively. The parametrizations obtained from the simultaneous blast-wave fits are shown as solid lines. The  deviation of the $p_{\rm T}$ 
spectrum from the blas-wave model is shown in the bottom panels, indicating the fitting range using colored markers.}}
\label{fig:bwEPOS3}
\end{figure*}

Shown in Fig.~\ref{fig:bwEPOS3} the results obtained from the blast-wave analysis of the $p_{\rm T}$ spectra simulated with \textsc{Epos}~3. In general, the observed effects are quite similar to those shown for \textsc{Pythia}~8 (Fig.~\ref{fig:bwPythia8}), 
even though the statistics is notably limited for jets with higher $p_{\rm T}^{\rm jet}$. In low-multiplicity events with no jets produced (Fig.~\ref{fig:bwEPOS3}(a) left column) charged kaons show worse model description than those seen for \textsc{Pythia}~8 
(Fig.~\ref{fig:bwPythia8}(a) left column) which are similar in shape and magnitude to (anti)protons. Also, the corresponding $\chi^{2}$ values are remarkably higher in events without jets simulated with (without) hydrodynamical evolution, i.e. 
$\chi^{2}/\mathrm{n.d.f} = 5.55$ ($9.18$), than in those simulated with (and without) color reconnection in \textsc{Pythia}~8.\\
Low-$z$ $p_{\rm T}$ spectra in \textsc{Epos}~3 for all particle species are better described by the blast-wave model. It is noteworthy that it is hard to make strong conclusion for the highest $p_{\rm T}^{\rm jet}$ event class due to its limited statistics, but 
the points tend to show similar behavior compared to the former event class with jets $5<p_{\rm T}^{\rm jet}<10$\,GeV/$c$.\\
In high-$z$ events, the model description of the $p_{\rm T}$ spectra show the same behavior seen in \textsc{Pythia}~8. Namely, with the inclusion of hydrodynamical component in the simulation, the agreement is somewhat better for the case with the hydro option. 
Besides that, there is a clear dependence seen between the two options (shown with full and open markers) also for charged pions and kaons. The $p_{\rm T}$ spectra have a weak $p_{\rm T}^{\rm jet}$-dependent behavior and their agreement with the blast-wave model 
improves towards higher $p_{\rm T}^{\rm jet}$, with the change of $\chi^{2}/\mathrm{n.d.f}$ from  $1.33$ ($6.74$) to $0.77$ ($1.29$) as we consider events with (without) setting hydro option, respectively.

To quantify the importance of jets in events where flow patterns are generated with hydrodynamics or color reconnection, Fig.~\ref{fig:TkinBeta} shows the correlation between the blast-wave parameters $T_{\rm kin}$ and $\langle \beta_{\rm T} \rangle$. 
Results are shown for different $z$ multiplicity classes which are indicated by different marker sizes and increase from low  to high multiplicity. Besides the multiplicity selection, also shown the case when we consider the selection on the hardness of the 
event by imposing a cut on $p_{\rm T}^{\rm jet}$.\\
In continuation to the previous observation based on the fit qualities for various event classes in \textsc{Pythia}~8 and in \textsc{Epos}~3, by looking at the evolution of the blast-wave fit parameters in terms of $z$ and $p_{\rm T}^{\rm jet}$ we see that:

\begin{figure*}
  \includegraphics[width=0.93\textwidth]{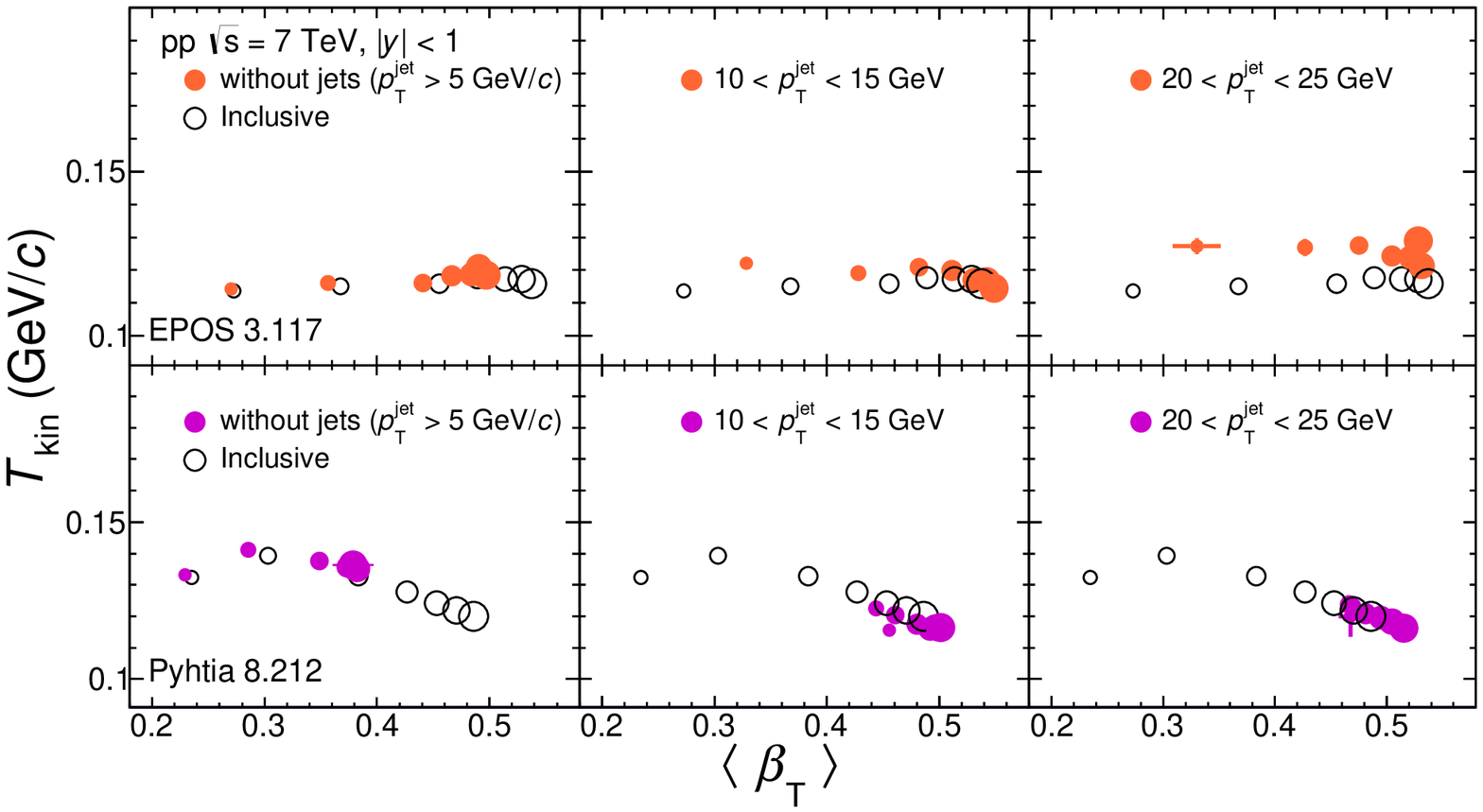}     
\caption{(Color online) \small{Correlation between two fit parameters obtained from the blast-wave analysis, the kinetic temperature ($T_{\rm kin}$) and the 
average transverse expansion velocity ($\langle \beta_{\rm T} \rangle$) of the system. Results for pp collisions at $\sqrt{s}=$ 7\,TeV simulated with \textsc{Epos}~3 
and \textsc{Pythia}~8 are presented in the top and the bottom rows, respectively. The size of the markers increases with the event multiplicity. Results with 
(full markers) and without (empty markers) a selection on $p_{\rm T}^{\rm jet}$ are compared.}}
\label{fig:TkinBeta}
\end{figure*}

\begin{itemize}
\item For events containing jets and being in the same multiplicity class (indicated by the same marker size), $\langle \beta_{\rm T} \rangle$ increases with respect to the case without any selection on the hardness (inclusive case indicated by open markers). 
This is a somewhat natural consequence of the auto-correlation bias due to hard partons inducing a large boost. Moreover, by looking at, the case of jets with $20<p_{\rm T}^{\rm jet}<25$\,GeV/$c$ and the highest multiplicity class ($5<z<6$), the effect is weaker 
in \textsc{Epos}~3 ($\approx$~0.6\,\%) than in \textsc{Pythia}~8 ($\approx$~6.8\,\%).  Implicitly, this is also illustrated in the smaller multiplicity dependence of $\langle \beta_{\rm T} \rangle$ obtained in \textsc{Pythia}~8 than that observed in \textsc{Epos}~3 
where the role of jets is considerably weaker.
\item In low-$z$ events (as explained earlier)  the $p_{\rm T}$ spectra deviate from the blast-wave model\,--\,in constrast to events with higher $z$\,--\,and the agreement improves with increasing $p_{\rm T}^{\rm jet}$. In connection to those observations, taking 
low-$z$ parameters in Fig.~\ref{fig:TkinBeta} we see a weak $p_{\rm T}^{\rm jet}$ dependence as a function of $\langle \beta_{\rm T} \rangle$ in \textsc{Epos}~3 whereas in \textsc{Pythia}~8 for a given low-$z$ class events with increasing $p_{\rm T}^{\rm jet}$ 
experience a larger radial flow velocity $\langle \beta_{\rm T} \rangle$. In addition, one sees that in \textsc{Pythia}~8 the larger the $\langle \beta_{\rm T} \rangle$ the smaller the $T_{\rm kin}$ reaching at slightly smaller values for the highest $z$ class than 
those in \textsc{Epos}~3 where the $T_{\rm kin}$ dependence of $\langle \beta_{\rm T} \rangle$ is almost flat.
\end{itemize}

  
\section{Conclusions}

In this work, we have presented a study using two event generators, \textsc{Epos}~3 and \textsc{Pythia}~8, exploring an observable which is aimed at ruling out or validating the underlying physics mechanism (hydrodynamics or color reconnection) generating 
radial flow patterns in pp collisions. Specifically, we exploit the fact that, by construction, color reconnection produces a strong coupling between the hard (hard partons) and soft (soft and semi-hard partons) components of the interaction. To this end, we 
have studied the $p_{\rm T}$ spectra of charged pions, kaons and (anti)protons as a function of the event multiplicity and the transverse momentum of the leading jet. Our main findings can be summarized as follows:

\begin{itemize}

\item The multiplicity dependence of the proton-to-pion 
particle ratios show a depletion for $p_{\rm T}\lesssim 1-2$\,GeV/$c$ and an enhancement above, ending up with a bump at around $p_{\rm T}\approx 3$\,GeV/$c$. Apart from the multiplicity selection, a more differential classification was done by 
using the leading jet transverse momentum ($p_{\rm T}^{\rm jet}$). At extremely low multiplicity, it is possible to find a subclass of events where radial flow patterns arise\,--\,despite the fact that at very low multiplicity\,--\,hydrodynamics 
cannot be applied and color reconnection effects are small. This feature is present both for \textsc{Pythia}~8 and \textsc{Epos}~3.

\item In high-multiplicity events  the particle composition is very different in \textsc{Pythia}~8 and \textsc{Epos}~3 which is nicely visible on the proton-to-pion ratios when multiplicity and $p_{\rm T}^{\rm jet}$ vary. In \textsc{Epos}~3 the magnitude 
of the  bump significantly increases with decreasing $p_{\rm T}^{\rm jet}$. On the contrary, in \textsc{Pythia}~8 the height of the peak does not change with $p_{\rm T}^{\rm jet}$.

\item Furthermore, the agreement between the blast-wave  model and the charged pion, kaon and (anti)proton spectra  in events classified using multiplicity and $p_{\rm T}^{\rm jet}$,  significantly improves with the increase of the leading 
jet $p_{\rm T}$. More importantly, this agreement was found to be the best  in low-multiplicity events having jets, suggesting the presence of the collective-like behavior caused by jets.

\item The multiplicity  and jet transverse momentum dependence of the blast-wave fit parameters, i.e. the average transverse expansion velocity ($\langle \beta_{\rm T} \rangle$) and the kinetic temperature ($T_{\rm kin}$), also shows effects which can be 
used to discriminate between different models. It is found that the multiplicity dependence of $\langle \beta_{\rm T} \rangle$ is more affected by jets in \textsc{Pythia}~8 than in \textsc{Epos}~3. 
\end{itemize}

  
\ack

We acknowledge Gergely G\'abor Barnaf{\"o}ldi, Peter Christiansen, Eleazar Cuautle, Arturo Fern\'andez and Guy Pai\'c for the critical reading 
of the manuscript and the valuable discussion and suggestions. We also acknowledge Klaus Werner for allowing us the usage of EPOS 3.117 and for 
the useful instructions. 

Support for this work has been received from CONACYT under the grant No. 260440; from DGAPA-UNAM under PAPIIT grant IA102515. In addition, this 
work was supported by Hungarian OTKA grants NK106119, K120660 and NIH TET 12 CN-1-2012-0016.


\end{document}